# How fractional counting of citations affects the Impact Factor: Normalization in terms of differences in citation potentials among fields of science

Loet Leydesdorff

Amsterdam School of Communications Research (ASCoR), University of Amsterdam, Kloveniersburgwal 48, 1012 CX Amsterdam, The Netherlands; loet@leydesdorff.net; http://www.leydesdorff.net

&

Lutz Bornmann

Max Planck Society, Office of Research Analysis and Foresight, Hofgartenstraße 8, 80539 Munich, Germany; bornmann@gv.mpg.de.

**Abstract**

The ISI-Impact Factors suffer from a number of drawbacks, among them the statistics—why should one use the mean and not the median?—and the incomparability among fields of science because of systematic differences in citation behavior among fields. Can these drawbacks be counteracted by counting citation weights fractionally instead of using whole numbers in the numerators? (*i*) Fractional citation counts are normalized in terms of the citing sources and thus would take into account differences in citation behavior among fields of science. (*ii*) Differences in the resulting distributions can be tested statistically for their significance at different levels of aggregation. (*iii*) Fractional counting can be generalized to any document set including journals or groups of journals, and thus the significance of differences among both small and large sets can be tested. A list of fractionally counted Impact Factors for 2008 is available online at http://www.leydesdorff.net/weighted_if/weighted_if.xls. The in-between group variance among the thirteen fields of science identified in the U.S. *Science and Engineering Indicators* is not statistically significant after this normalization. Although citation behavior differs largely between disciplines, the reflection of these differences in fractionally counted citation distributions could not be used as a reliable instrument for the classification.

## Introduction

This study has three objectives:

1. In a previous communication, Leydesdorff & Opthof (2010a) proposed using fractional counting of citations as a means to normalize impact factors in terms of differences in citing behavior ("citation potential") among disciplines. We apply this normalization to the 6,598 journals included in the *Journal Citation Reports* 2008 (Science Edition) and compare the results with the ISI Impact Factors.
2. Using the thirteen fields identified by ipIQ for the purpose of developing the *Science and Engineering Indicators 2010* (NSB, 2010, at p. 5-30 and Appendix Table 5-24), it can be shown that this normalization by fractional counting reduces the in-between group variance in the impact factors by 81% (when compared with integer counting) and makes the remaining differences statistically not significant.
3. Because fractionally counted impact factors can be compared across fields, differences among the distributions in the numerators (that is, the fractions) can be tested statistically to determine if they can be used for classification among fields of science. For example, citation patterns in molecular biology are very different from citation patterns in mathematics. However, this classification is unreliable; other sources of variance, such as differences in publication behavior, cited half-life times, document types, etc., disturb classification on this basis.

For reasons of presentation, we discuss the third question before the second one in the results section. Refinements based on the discussion of field differences can then be tested as an additional model (Model 4) when answering the second question above.

Let us first turn to the theoretical relevance of these questions. The well-known impact factor (IF) of the Institute of Scientific Information (ISI)—presently owned by Thomson Reuters—is defined as the average number of references to each journal in a current year to “citable items” published in that journal during the two preceding years. Ever since its invention in 1965 (Sher & Garfield, 1965; Garfield, 1972 and 1979a), this ISI-IF has been criticized for a number of seemingly arbitrary decisions involved in its construction. The possible definitions of “citable items”—articles, proceedings papers, reviews, and letters—the choice of the mean (despite the well-known skew in citation distributions; Seglen, 1992), the focus on two preceding years as representation of impact at the research front (Bensman, 2007), etc., have all been discussed in the literature, and many possible modifications and improvements have been suggested (recently, e.g., Althouse *et al*., 2009).

In response, Thomson Reuters has added the five-year impact factor (ISI-IF-5), the Eigenfactor Score, and the Article Influence Score (Bergstrom, 2007; Rosvall & Bergstrom, 2008) to the journals in the online version of the *Journal Citation Reports* (*JCR*) in 2007. Most recently, the JCR 2009 also introduced a new measure of relatedness among journals (Pudovkin & Garfield, 2002). While the extension of the IF to a five-year time window is straightforward, the JCR interface at the Web of Science itself fails to explain the more recently added measures because they can perhaps be considered as too complex for library usage (Adler *et al*., 2009, at p. 12; Waltman & Van Eck, 2010a, at p. 1483; cf. West *et al*., 2008).

Two indicators among the set (e.g., Leydesdorff, 2009; Van Noorden, 2010) stand out for their intuitive ease of understanding: ISI-IF as an average number of citations in the current year to publications in the two preceding years, and the cumulative citations to each journal ("total cites") as an indicator of a journal's overall visibility (Bensman, 2007). "Total cites" includes the historical record of the journal and therefore can also be considered as an indicator of prestige—potentially to be defined differently from a reputation among specialists (Bollen *et al*., 2006; Brewer *et al*., 2001). *Science* and *Nature* are the best-known examples of multidisciplinary journals with high prestige. The influence of a prestigious journal may reach down all the way into specialties to the level of strategic interventions, such as the role played by *Science* in the emergence of nanotechnology around the year 2000 (Leydesdorff & Schank, 2008).

In other words, the citation networks among journals contain both a hierarchical stratification and a network structure in which different densities represent specialties which can be expected to operate in parallel. The resulting system therefore is complex and not fully decomposable (Simon, 1973). Some journals span the specific distance between two specialties, and this is often reflected in their titles (e.g., *Limnology and Oceanography*). Other journals span larger sets of specialties, such as the *Journal of the American Chemical Society* (*JACS*), which primarily relates organic, inorganic, and physical chemistry as major subject areas within chemistry, but also relates to other subdisciplinary structures such as biochemistry and electrochemistry (Bornmann *et al*., 2007; Leydesdorff & Bensman, 2006). The *Proceedings of the National Academy of Science of the USA* (*PNAS*), for example, can be compared with *Science* and *Nature* for its transdisciplinary role, but with the *JACS* for its role in recombining citations to specialties in the various areas of bio-medicine and molecular biology.

In summary, journals cannot easily be compared, and classification systems based on citation patterns hence tend to fail. A variety of perspectives remains possible; in different years, some perspectives may be more important than others. Indexes such as the ISI Subject Categories accommodate this multitude of perspectives by listing journals under different categories for the purpose of information retrieval. Information retrieval, however, provides an objective different from analytical distinctions (Pudovkin & Garfield, 2002, at p. 1113n.; Rafols & Leydesdorff, 2009).

Efforts to classify journals using multivariate statistics of citation matrices have been somewhat successful at the local level (Leydesdorff, 2006) and more recently also at the global level (Rosvall & Bergstrom, 2008 and 2010), but the positions of individual journals on the borders between specialties remain difficult to determine with precision. Thus, normalization of the ISI-IFs (or other impact indicators) using one classification of journals or another has hitherto remained an unsolved problem.

## Integer and fractional counting of citations

Most efforts to classify journals in terms of fields of science have focused on correlations between citation patterns in core groups assumed to represent scientific specialties. However, there may be other statistical patterns which are field specific and allow us to classify journals. Garfield (1979a and b), for example, proposed the term “citation potential” for systematic differences among fields of science based on the average number of references. For example, in the bio-medical fields long reference lists (for example, with more than 40 references) are common, but in mathematics short lists (with fewer than six references) are the standard. These differences are a consequence of differences in citation cultures among disciplines, but can be expected to lead to significant differences in the ISI-IFs among fields of science because the chance of being cited is systematically affected.

We propose to use fractional counting of citations as a means to normalize for these differences: using fractional counting, a citation in a *citing* paper containing *n* references counts for only $(1/n)^{th}$ of overall citations instead of a full point (as is the case with integer counting). The ISI-IF is based on integer counting; this IF is thus sensitive to differences in citation behavior among fields of science. A fractionally counted IF would correct for these differences in terms of the sources of the citations. Such normalization therefore can also be called “source-normalization” (e.g., Moed, 2010; Van Raan *et al.*, 2010; Waltman & Van Eck, 2010b; Zitt, 2010).

The suggestion to use fractional counting to solve the problem of field-specific differences in citation impact indicators originated from a discussion of measurement issues in institutional research evaluation (Opthof & Leydesdorff, 2010; Van Raan *et al*., 2010; Leydesdorff & Opthof, 2010b). Institutes are populated with scholars with different disciplinary backgrounds, and research institutes often have among their missions the objective to integrate bodies of knowledge "interdisciplinarily" (Wagner *et al*., 2009; Leydesdorff & Rafols, 2010). In such a case, one is confronted with the need to normalize across fields of science because citation practices differ widely across the disciplines and even within them among specialty areas. Resorting to the ISI Subject Categories for normalization would beg the question in such cases. Interdisciplinary work may easily suffer in the evaluation from being misplaced in a categorical classification system (Laudel & Orrigi, 2006).

The use of fractional counting in citation analysis provides us with a tool to normalize in terms of the citation behavior of the *citing* authors in a current year.[1] Fractional counting of the citations can be expected to solve the problem of normalization among different citation practices because each unique citation is positioned relatively to the citation practice of the author(s) of the citing document (Bornmann & Daniel, 2008; Leydesdorff & Amsterdamska, 1990). Otherwise, comparing these uneven units in an evaluation, one might erroneously conclude that a university could improve its position in the citation ranking by closing its mathematics department or that a publishing house would be able to improve the impact of its journals by cutting the set at the lower end of the distribution of ISI-IFs.

[1] An approach to normalize citation impact at the field level from the *cited* side was recently proposed by Stringer *et al*. (2010).

Furthermore, Garfield (2006) noted that larger journals can be expected to serve larger communities, and therefore there is no *a priori* reason to expect them to have higher ISI-IFs. Althouse *et al*. (2009) distinguished between two sources of variance: differences between fields are caused mainly by differences in the ratio of references to journals included in the JCR set—as opposed to references to so-called "non-source items" (e.g., books)—whereas differences in the lengths of reference lists are mainly responsible for inflation in the ISI-IFs over time.

The application of the tool of fractional counting of citations to journal evaluation was anticipated by Zitt & Small (2008) and Moed (2010). Zitt & Small (2008) proposed the Audience Factor (AF) as another indicator, but used the *mean* of the fractionally counted citations to a journal (Zitt, 2010). This mean then was divided by the mean of all journals included in the *SCI*. Unlike a mean (or a median, range, or variance), however, a ratio of two means no longer contains a statistical uncertainty. The differences between these ratios, therefore, cannot be tested for their significance, and error in the measurement can no longer be specified.

In a similar vein, Moed (2010) divided a modified IF (with a window of three years and a somewhat different definition of citable issues) by the *median* of the citation potentials in the Scopus database. He proposed the resulting ratio as the Source Normalized Impact per Paper (SNIP) which is now in use as an alternative to the IF in the Scopus database (Leydesdorff & Opthof, 2010a). Note that the IF itself can be considered as a mean and therefore a proper statistic; the underlying distributions of IFs can be compared using standard tests (e.g., Kruskal-Wallis or ANOVA; cf. Bornmann, 2010; Opthof & Leydesdorff, 2010; Plomp, 1992; Pudovkin & Garfield, in print; Stringer *et al*., 2010).

In summary, the distributions of citations in the citing documents can be compared in terms of means, medians, variances, and other statistics. Differences among document sets can be tested for their significance independently of whether one uses journals, research groups, or other aggregating variables for the initial delineation of document sets. Although this can be done equally for fractional and integer counting, our hypothesis is that the difference between these two counting methods for citations is caused by the variation in citation behavior among fields.

Unlike the ISI-IF, one can expect that the distributions resulting from fractional counting of the citations will be comparable among fields of science. As a second objective, we will test whether one can use the differences in the distributions for distinguishing among journal sets in terms of fields of science. Leydesdorff & Opthof (2010a) developed the proposed method for the case of the five journals which were used by Moed (2010) for introducing the SNIP indicator. In this study, we first show that the quasi-IFs based on fractional counting enable us to distinguish mathematics journals from journals in molecular biology. However, this test fails at the finer-grained level of specialties and journals within fields of science. Citation behavior varies with fields of science, but not among specialties within fields.

## Methods and materials

### *Data processing*

Data was harvested from the CD-Rom versions of the *SCI* 2008 and the *JCR* 2008. Note that the CD-Rom version of the *SCI* covers fewer journals than the *Science Citation Index-Expanded* (*SCI-E*) that is available at the Web of Science (WoS; cf. Testa, 2010). (This core set is also used for the *Science and Engineering Indicators* of the National Science Board of the USA.[2]) The data on the CD-Rom for 2008 contains 1,030,594 documents published in 3,853 journals.[3] Of these documents, 944,533 (91.6%) contain 24,865,358 cited references. Each record in the ISI set contains conveniently also the total number of references ($n$) at the document level. Each citation can thus be weighted as $1/n$ in accordance with this number in the citing paper.

In a first step, the references to the same journal within a single citing document were aggregated. For example, if the same document cites two articles from *Nature*, the fractional citation count in this case is $2/n$. In this step, citations without a full publication year (e.g., "in press") were no longer included. This aggregation led to a file with 14,367,745 journal citations; 9,702,753 of these (67.5%) contain abbreviated journal names that we were able to match with the abbreviated journal names in the list of 6,598 journals included in the *SCI-E* in 2008.[4]

[2] Ken Hamilton, communication at the email list sigmetrics@listserv.utk.edu, 3 May 2010.
[3] We found 3,853 journal titles in the download. Ken Hamilton (personal communication, June 1, 2008) reports 3,737 journals used for preparing the *Science and Egineering Indicators 2010* (NSB, 2010) based on the same files (2008).

There was no *a priori* reason to limit our exercise to the smaller list of the CD-Rom version of the *SCI* because all journals can be cited and IFs for all (6,598) journals in the *SCI-E* are available for the comparison. However, one should keep in mind that only citations provided by the 3,853 journals in the smaller set (of the *SCI*) are counted in this study given the database that is used as source data on the citing side. Thus, one can expect significantly lower numbers of references than those retrievable at the WoS.

A match in terms of the journal abbreviations in the reference list was obtained in 6,566 (99.5%) of the 6,598 *JCR*-journals. These 6,566 journals contain 19,200,966 (77.2%) of the total of 24,865,358 original references. The citation numbers in this selection are used for computing the total cites for each journal, both fractionally and as integer numbers. When counted fractionally the number of references is 555,510.07 (that is, 2.89% of the total number of references or, in other words, with an average of 34.6 references per citing article).

[4] As an exception, the journal name 'Arthritis and Rheumatism' is abbreviated with 'Arth Rheum/Ar C Res' in the journal list, but with 'Arth Rheum' when used in cited references.

| SCI 2008 | Citations to all years | Citations to 2006 and 2007 |
|---|---|---|
| *Nr of cited references* | 24,865,358 | 3,898,851 |
| *Nr of abbreviated journal titles* | 14,367,745 | 2,936,157 |
| *Nr of abbreviated journal titles matching* | 9,702,753 | 2,422,430 |
| *Nr of cited references after matching* | 19,200,966 | 3,320,894 |
| *Nr of cited references fractionally counted* | 555,510.07 | 596,755.99 (103,828.70) |
| *Average nr of references/paper* | 34.6 | 5.6 |

**Table 1**: Descriptive statistics of the citation data 2008 and the various steps in the processing.

By setting a filter to the citations from 2006 and 2007 in the original download, the numerators of the weighted quasi-IFs can be calculated from the same 25M references; the same procedure was repeated for this subset. The third column of Table 1 shows the corresponding numbers.

The 2008-file contains 3,898,851 references to publications with 2006 or 2007 as publication years (in 187,966 journals). When counted fractionally, this number is 124,946.59 citations. 103,828.70 (83.1%) of this count is included in the analysis using the 6,566 journals for which the journal abbreviations in the reference lists could be matched with the full journal names listed in the JCR. However, when divided by the (much smaller) number of cited references from only the two previous years, the average number of citations per document is 5.6 and the fractional count adds up across these journals to 595,755.99. We use this latter normalization below because it corresponds, in our opinion, to the intended focus of the IF on citations at the research front (that is, the last two years).

For the denominator of our quasi-IFs, we used the sum of the numbers of citable issues in 2006 and 2007 as provided by the *JCR*s of these respective years. By setting a filter to the period 2003-2007, one could analogously generate a five-year IF, both weighted or without weighting. However, we limit the discussion here to the two-year IF and follow strictly the definitions of the ISI (Garfield, 1972). Of the 6,598 journals listed in the JCR-2008 only 5,794 could thus be provided with a value for the denominator of the IF in 2008 based on values for the number of citable items in the two preceding years larger than zero. In a next step, we use exclusively the references provided to the 2006 and 2007 volumes of the 5,742 journals which have both a non-zero value in the numerator (2008) and in both terms of the denominator (2006 and 2007, respectively). These 5,742 journals contain 3,255,133 references or fractionally counted 583,833.98 references, to publications in 2006 and 2007.

*Testing for between-group variances among fields of science*

We will test the extent to which the normalization implied by using fractional counting reduces the between-group variance in relation to the within-group variance for the case of the thirteen fields of science identified by ipIQ for the purpose of developing the *Science and Engineering Indicators 2010* (NSB, 2010, at p. 5-30 and Appendix Table 5-24). We chose this classification because it is reflexively shaped and regularly updated on a journal by journal basis without automatic processing. Furthermore, journals are uniquely attributed to a broad field. However, the attribution is made only for the approximately 3900 journals used as original source data in both this study and the *Science and Engineering Indicators* of the NSF.

A two-level regression model will be estimated in which the quasi-IFs of journals are level-1 units and the 13 fields are level-2 clusters. Various two-level regression models are possible—depending on the scale of the dependent variable (here: quasi-IFs). Since IFs for journals are based on citation counts for the papers published in these journals, citations can be considered as count data. In the case of count data, a Poisson distribution is the best assumption (Cameron & Trivedi, 1998). Thus, we shall calculate a two-level random-intercept Poisson model. In order to handle overdispersion at level 1 (measured as large differences between the mean and the variance of the IFs) in this model, we follow Rabe-Hesketh & Skrondal's (2008) recommendation to use the sandwich estimator for the standard errors.[5]

*Using differences in citation behavior for the classification*

The fractional counts of the citations provide us with distributions indicating citation behavior at the level of each journal. Which statistics could be useful to test these multiple citation distributions of different sizes for the significance of their homogeneity and/or differences?

Let us first note that in the case of integer counting and aggregated journal citations, one can expect the distributions in homogenous sets to be highly skewed (Leydesdorff &

[5] We calculated also a normal regression analysis after lognormalizing the dependent variable in order to receive the between-field variance. This procedure provides results that have the same tendency.

Bensman, 2006; Seglen, 1992; Stringer *et al*., 2010). This expectation is likely to hold also for fractional counting. Before using parametric statistics for highly skewed data (e.g., ANOVA) a log-normalising transformation is recommended (Allison, 1980). However, we did not log-normalize the data because our objective is to test the effects of fractional counting on the field effect in the IF. This field effect may partly disappear by log-normalizing the data albeit it less so for the two-year citation window used for the IF (Stringer *et al*., 2010). Thus, we would be at risk of confounding two different research questions.

Post hoc pairwise comparisons can be performed after obtaining a significant omnibus F with ANOVA. Among the post hoc tests which are available in SPSS for multiple comparisons, one may prefer to choose one of the tests which do not *ex ante* assume equal variance (for example, Dunnett's C test). However, this assumption about homogeneity in the variance itself can first be tested using Levene's Test for Equality of Variances (available within ANOVA). If alternatively the assumption holds, one can use the Tukey test which—as implemented in SPSS—includes controls for testing the significance of the differences among *multiple* samples.

### A note about differences with the ISI-IFs

The ISI-IFs are produced by the team at Thomson Reuters responsible for the *JCR.* The sum of the total number of times the 6,598 journals included in the *Journal Citations Report* 2008 (for the *SCI*) are cited, is 29,480,301. This is 53.5% more than the total number of citations (19,200,966) to these journals retrieved above (Table 1). Unlike our download, the *JCR* is based on publication years.

Furthermore, Thomson Reuters has hitherto followed a procedure for generating the JCR that is uncoupled from the production of the CD-Rom version of the *SCI*. Like the WoS, the *JCR* is based on the *SCI-E* that includes many more (citing) journals than the CD-Rom version of the *SCI.* While the JCR contained 6,598 journals in 2008, the CD-Rom version contained only 3,853 source journals: these 58.4% of the journals, however, cover 65.1% of the cited references (cf. Testa, 2010).[6]

The CD-Rom versions are based on processing dates between January 1 and December 31 while the JCR is based on publication years, but on the basis of a decision in each year to produce the database at a cut-off date in March.[7] In both these databases, the publication years are thus incomplete, and therefore cannot be expected to correspond to the numbers retrievable from the online version of the WoS (McVeigh, *personal communication*, April 7, 2010). Furthermore, journals may be added to the WoS version which are backtracked to previous years—and can thus be retrieved online—while both the JCR and the CD-Rom versions can no longer be changed after their production. Thus, the various versions of the *SCI* cannot directly be compared. In the meantime, there is a blossoming literature complaining about the impossibility of replicating journal IFs using the WoS (e.g., Brumback, 2008a and b; Rossner *et al.*, 2007 and 2008; Pringle, 2008).[8]

---

[6] http://thomsonreuters.com/products_services/science/science_products/a-z/science_citation_index [Accessed on May 24, 2010].
[7] The WoS allows for searching with publication dates or calendar years.

**Results**

Let us nevertheless and as a first control compare the ISI-IFs as provided by the JCR 2008 with the quasi-IFs retrieved from the CD-Rom version of the *SCI 2008*. Table 2 provides the Pearson and Spearman rank correlations between the ISI-IF, the quasi-IF derived from the download of 2008, and the corresponding quasi-IF based on fractional counting. Not surprisingly—because of the high value of *N*—all correlations are significant at the 0.01 level. In the rightmost column, we also added the fractionated citations/publications ratio for 2008, for reasons to be explained below.

| | ISI-IF | Quasi-IF (integer) | Quasi-IF (fractional) | Fractional c/p 2008 |
|---|---|---|---|---|
| ISI-IF | | .898(**)<br>5742 | .835(**)<br>5742 | .669(**)<br>5687 |
| Quasi-IF (integer) | .971(**)<br>5742 | | .937(**)<br>5742 | .770(**)<br>5687 |
| Quasi-IF (fractional) | .926(**)<br>5742 | .937(**)<br>5742 | | .813(**)<br>5687 |
| Fractional c/p 2008 | .746(**)<br>5687 | .771(**)<br>5687 | .818(**)<br>5687 | |

** Correlation is significant at the 0.01 level (2-tailed).

**Table 2**: Correlations between the ISI-IF, quasi-IFs based on integer and fractional counting, and fractionally counted citations divided by publications in 2008.[9] The lower triangle provides the Pearson correlations ($r$) and the upper triangle the corresponding Spearman rank-order correlations ($\rho$).[10]

[8] We acknowledge Roger A. Brumback for reporting these references after a literature search at the list sigmetrics@listserv.utk.edu, November 7, 2008.

[9] Of these 5,742 journals, 55 journals did not contain a number of issues in the *JCR* 2008. (Of the 6,598 journals contained in the *JCR* 2008, 133 were not attributed a number of issues.)

[10] Using the Kolmogorov-Smirnov test, it could be inferred that the distributions for all four variables cannot be assumed to follow a normal distribution.

As could be expected, the quasi-IF based on integer counting correlates higher with the ISI-IF than the one based on fractional counting. These correlations confirm that our quasi-IFs can be considered similar to the ISI-IF in nature, although there may be important differences at lower levels of aggregation. For example, the Pearson correlation ($r$) between the distributions of fractional and integer counts is only 0.464 ($\rho = 0.654$;[11] $p < 0.01$) for the 9,702,753 references matching in the total set, and $r = 0.128$ ($\rho = 0.261$;[13] $p < 0.01$) for the 2,422,430 references to publications only in 2006 and 2007. The two IFs (based on integer and fractional counting, respectively) are very different in terms of the numerators of the IFs. Yet, the quasi-IF based on fractional counting can explain more than 85% of the variance in the ISI-IF ($r^2 = (.926)^2 = .857$).

**Is field normalization accomplished by fractional counting?**

Since it is not possible to test the 5,742 journals against one another using multiple comparisons in SPSS, we first focused on the five journals which were discussed by Moed (2010) and Leydesdorff & Opthof (2010a). (In these previous studies different criteria were used for reason of comparison with the SNIP indicator of *Scopus*.) Table 3 teaches us that the rank-order of the quasi-IFs among these five journals is different when counted fractionally instead of using integer counting: *Annals of Mathematics* in this case has a value (1.416) higher than that of *Molecular Cell* (1.143), while the ISI-IF and the quasi-IF based on integer counting show the expected (large) effect of differences among the two corresponding fields of science. This provides us with a first indication that our method for the correction of citation potentials might work.

[11] The Spearman correlations are estimates based on sampling (within SPSS) because of the large number of cases.

| *2008* | ISI-IF2 | IF (integer) | IF (fractional) | IF (fractional)* |
|---|---|---|---|---|
| 1. *Invent Math* | 2.287 | 1.294 | 0.595 | 0.064 |
| 2. *Mol Cell* | 12.903 | 11.011 | 1.143 | 0.247 |
| 3. *J Electron Mater* | 1.283 | 0.868 | 0.255 | 0.043 |
| 4. *Math Res Lett* | 0.524 | 0.323 | 0.175 | 0.016 |
| 5. *Ann Math* | 3.447 | 2.688 | **1.416** | 0.129 |

**Table 3**: ISI-IF and quasi IF for integer and fractional counting.
* The right-most column additionally provides the IF based on fractional counting, but using all references for the normalization.

We added to Table 3 a right-most column with the values of the IF based on fractional counting, but using the total number of citations (and not only the ones to publications in 2006 and 2007) for the normalization. These values are much smaller because of the larger numbers in the respective denominators of the fractions, and—as perhaps to be expected—they show the effects of fractional counting to a smaller extent. In other words, the interesting difference in the rank order is generated by using fractional counting exclusively on the basis of references to publications in 2006 and 2007. We therefore use this latter normalization in the remainder of this study.

The Levene test for the homogeneity of variances teaches us that these five journals are significantly different and that thus a test which is not based on this assumption should be used. As noted, we use Dunnett's C-test in such cases.

| (I) jnr | (J) jnr | Mean Difference (I-J) | Std. Error | 95% Confidence Interval | |
|---|---|---|---|---|---|
| | | Lower Bound | Upper Bound | Upper Bound | Lower Bound |
| **1** | 2 | .351876209(*) | .024918239 | .28313736 | .42061506 |
| | 3 | .122373032(*) | .029362612 | .04152074 | .20322533 |
| | **4** | **-.054643614** | .048119007 | -.18982705 | .08053982 |
| | **5** | **-.071077267** | .033497287 | -.16341644 | .02126191 |
| 2 | 1 | -.351876209(*) | .024918239 | -.42061506 | -.28313736 |
| | 3 | -.229503177(*) | .015755897 | -.27268039 | -.18632596 |
| | 4 | -.406519823(*) | .041249535 | -.52315517 | -.28988448 |
| | 5 | -.422953476(*) | .022542261 | -.48503123 | -.36087572 |
| 3 | 1 | -.122373032(*) | .029362612 | -.20322533 | -.04152074 |
| | 2 | .229503177(*) | .015755897 | .18632596 | .27268039 |
| | 4 | -.177016646(*) | .044076847 | -.30117111 | -.05286219 |
| | 5 | -.193450300(*) | .027375132 | -.26872141 | -.11817919 |
| **4** | **1** | **.054643614** | .048119007 | -.08053982 | .18982705 |
| | 2 | .406519823(*) | .041249535 | .28988448 | .52315517 |
| | 3 | .177016646(*) | .044076847 | .05286219 | .30117111 |
| | **5** | **-.016433653** | .046932651 | -.14835491 | .11548761 |
| **5** | **1** | **.071077267** | .033497287 | -.02126191 | .16341644 |
| | 2 | .422953476(*) | .022542261 | .36087572 | .48503123 |
| | 3 | .193450300(*) | .027375132 | .11817919 | .26872141 |
| | **4** | **.016433653** | .046932651 | -.11548761 | .14835491 |

* The mean difference is significant at the .05 level.

**Table 4**: Multiple comparisons among the distributions of the fractional citation counts of the five journals listed in Table 3; Dunnett's C-test (SPSS, v15); no homogeneity in the variance assumed.

Table 4 shows that the fractional citation counts for the three mathematics journals (numbers one, four, and five) are *not* significantly different in terms of this test, while they are significantly different from the two non-mathematics journals (the numbers two and three) which additionally are significantly different from each other. Can this test for homogeneity in this proxy of citation behavior be used for the grouping of journals more generally?

## Testing significant differences in larger sets

ANOVA post-estimation pairwise comparison (SPSS, v. 15) allows for testing 50 cases at a time. How to select 50 from among the 5,742 journals in our domain? Most ISI Subject Categories contain more than 50 journals, but fortunately, the most problematic one of "multidisciplinary" journals contains only 42 journals. Preliminary testing of the fractional citation distributions of this set provided us with both counter-intuitive and intuitively expectable results. However, we saw no obvious way of validating the quality of the distinctions suggested by using Dunnett's C-test within this set.

Thus, we devised another test extending and generalizing from the above noted difference between the three mathematics journals and the two other journals. Can journals in mathematics and cellular biology (including *Molecular Cell*) be sorted separately using this method? For this purpose we used the 20 journals with highest ISI-IFs in the ISI Category *Mathematics*[12] and the 20 journals with highest ISI-IFs in the category of *Cell Biology*.[13]

In 2008, the top-20 mathematics journals range in terms of their ISI-IFs from 1.242 for *Communications in Partial Differential Equations* to 3.806 for *Communications on Pure and Applied Mathematics. Annals of Mathematics* and *Inventiones Mathematicae* are part of this set, but *Mathematical Research Letters* (with an ISI-IF of 0.524) is not.

[12] The ISI Category *Mathematics* contains 214 journal names with ISI-IFs ranging from zero to 3.806 for *Communications on Pure and Applied Mathematics.*

[13] The ISI Category *Cell Biology* contains 157 journal names with ISI-IFs ranging 0.262 for *Biologischeskie Membrany* to 35.423 for *Nature Reviews of Molecular Cell Biology.* (No ISI-IF 2008 is provided for *Animal Cells and Systems.*)

*Molecular Cell* is classified by Thomson Reuters both as *Biochemistry and Molecular Biology* [14] and *Cell Biology*. The top-20 journals in the latter category range in terms of their ISI-IF 2008 from 7.791 for the journal *Aging Cell* to 35.423 for *Nature Reviews of Molecular Cell Biology*. Thus, one can expect the two groups (*Mathematics* and *Cell Biology*) to be very different in terms of both their ISI-IFs—there is no overlap in the two ranges—and their citation practices. Table 5 provides the values for the ISI-IFs and our quasi-IFs—based on integer and fractional counting, respectively—for the two groups.

[14] This subject category contains 276 journal names with an ISI-IF ranging to 31.253 for *Cell*.

| Journal | ISI-IF 2008 | Quasi-IF (integer counting) | Quasi-IF (fractional counting) | fractionated c/p ratio 2008 | Journal | ISI-IF 2008 | Quasi-IF (integer counting) | Quasi-IF (fractional counting) | fractionated c/p ratio 2008 |
|---|---|---|---|---|---|---|---|---|---|
| *Commun Pur Appl Math* | 3.806 | 2.151 | 0.750 | 2.390 | *Nat Rev Mol Cell Bio* | 35.423 | 28.339 | 3.129 | 4.416 |
| *B Am Math Soc* | 3.500 | 1.667 | 0.575 | 3.909 | *Cell* | 31.253 | 25.226 | 2.499 | 7.354 |
| *Ann Math* | 3.447 | 2.688 | 1.416 | 4.794 | *Nat Med* | 27.553 | 20.669 | 2.284 | 6.156 |
| *J Am Math Soc* | 2.476 | 1.667 | 0.803 | 1.429 | *Annu Rev Cell Dev Bi* | 22.731 | 18.385 | 1.967 | 5.168 |
| *Mem Am Math Soc* | 2.367 | 1.469 | 0.729 | 1.313 | *Nat Cell Biol* | 17.774 | 14.392 | 1.408 | 2.829 |
| *Invent Math* | 2.287 | 1.294 | 0.595 | 2.543 | *Cell Stem Cell* | 16.826 | *n.a.* | *n.a.* [15] | 0.447 |
| *Acta Math Djursholm* | 2.143 | 1.526 | 0.748 | 3.201 | *Cell Metab* | 16.107 | 12.994 | 1.347 | 0.890 |
| *Found Comput Math* | 2.061 | 1.121 | 0.422 | 0.207 | *Gene Dev* | 13.623 | 10.684 | 1.015 | 2.759 |
| *Comput Complex* | 1.562 | 0.357 | 0.175 | 0.144 | *Trends Cell Biol* | 13.385 | 11.212 | 1.186 | 2.088 |
| *Duke Math J* | 1.494 | 0.924 | 0.465 | 1.412 | *Mol Cell* | 12.903 | 11.011 | 1.143 | 2.151 |
| *Publ Math Paris* | 1.462 | 0.273 | 0.098 | 0.103 | *Dev Cell* | 12.882 | 10.566 | 1.095 | 1.516 |
| *J Differ Equations* | 1.349 | 0.992 | 0.382 | 0.659 | *Curr Opin Cell Biol* | 12.543 | 10.266 | 1.018 | 2.259 |
| *Am J Math* | 1.316 | 0.789 | 0.406 | 1.881 | *Nat Struct Mol Biol* | 10.987 | 9.695 | 1.000 | 0.887 |
| *Constr Approx* | 1.308 | 0.723 | 0.281 | 0.439 | *Curr Opin Genet Dev* | 9.677 | 7.156 | 0.727 | 1.368 |
| *Nonlinear Anal Theor* | 1.295 | 0.540 | 0.217 | 0.179 | *Trends Mol Med* | 9.621 | 6.961 | 0.742 | 1.215 |
| *B Symb Log* | 1.294 | 0.618 | 0.422 | 0.263 | *Plant Cell* | 9.296 | 8.213 | 0.890 | 2.030 |
| *Adv Math* | 1.280 | 0.797 | 0.409 | 0.487 | *J Cell Biol* | 9.12 | 7.743 | 0.827 | 2.977 |
| *Random Struct Algor* | 1.253 | 0.663 | 0.310 | 0.444 | *Curr Opin Struc Biol* | 9.06 | 7.337 | 0.883 | 1.674 |
| *J Differ Geom* | 1.244 | 0.791 | 0.369 | 1.684 | *Embo J* | 8.295 | 7.055 | 0.769 | 4.390 |
| *Commun Part Diff Eq* | 1.242 | 0.856 | 0.307 | 0.648 | *Aging Cell* | 7.791 | 5.345 | 0.501 | 0.316 |
| Mean | 1.909 | 1.095 | 0.494 | 1.407 | Mean | 15.343 | 12.276 | 1.286 | 2.645 |
| Standard deviation | 0.835 | 0.612 | 0.295 | 1.356 | Standard deviation | 7.949 | 6.449 | 0.694 | 1.928 |

**Table 5**: IFs and quasi-IFs for the twenty journals with highest ISI-IFs in the ISI Subject Categories of *Mathematics* and *Cell Biology*.

[15] There is no number of issues listed for *Cell Stem Cell* in the *JCR* in 2006. This number is part of the denominator of an IF. However, the journal can be compared in terms of the citations provided in 2008 (that is, the numerator of the IF).

Table 5 shows that the mean of the quasi-IFs based on fractional counting remains more than twice as high for the 20 journals in molecular biology (1.286) than for the 20 journals in mathematics (0.494). Thus, the correction for the field level seems not complete. In an email communication (23 June 2010), Ludo Waltman suggested that the remaining difference might be caused by the different rates at which papers in the last two years are cited in these two fields. In the journals classified as *Cell Biology* almost all papers contain references to recent (that is in this context, the last two years) publications, while this is less than half of the papers in journals classified as *Mathematics*.[16]

On the basis of this reasoning, a citation window longer than two years would attenuate this remaining difference. For example, the IF-5 can be expected to do better for this correction than the IF-2. More radically, the accumulation of all citations—that is, "total cites"—divided by the number of publications (the c/p ratio) for all years would correct for the differences among journals in terms of their cited half-lives.[17] The right-most columns in each category of Table 5, however, show that a difference between the mathematics set and the cell-biology set remains even when fractionated c/p ratios—which include citations from all years—are used. Thus, the field-specific effects are further mitigated, but do not disappear. In other words, these differences cannot be fully explained by the citation potentials of the two different fields; the fields remain different.

[16] Waltman & Van Eck (2010b) therefore suggests an additional normalization based on the *average* number of references in the citing journal rather than straightforwardly using the citing publications as the reference standard.

[17] The assumption implied is that the fields grow proportionally in terms of the database. Since this is not likely, a shorter citation window may also have advantages.

Let us take a closer look into these differences and to the issue of whether we should include all or more previous years or only the last two years?

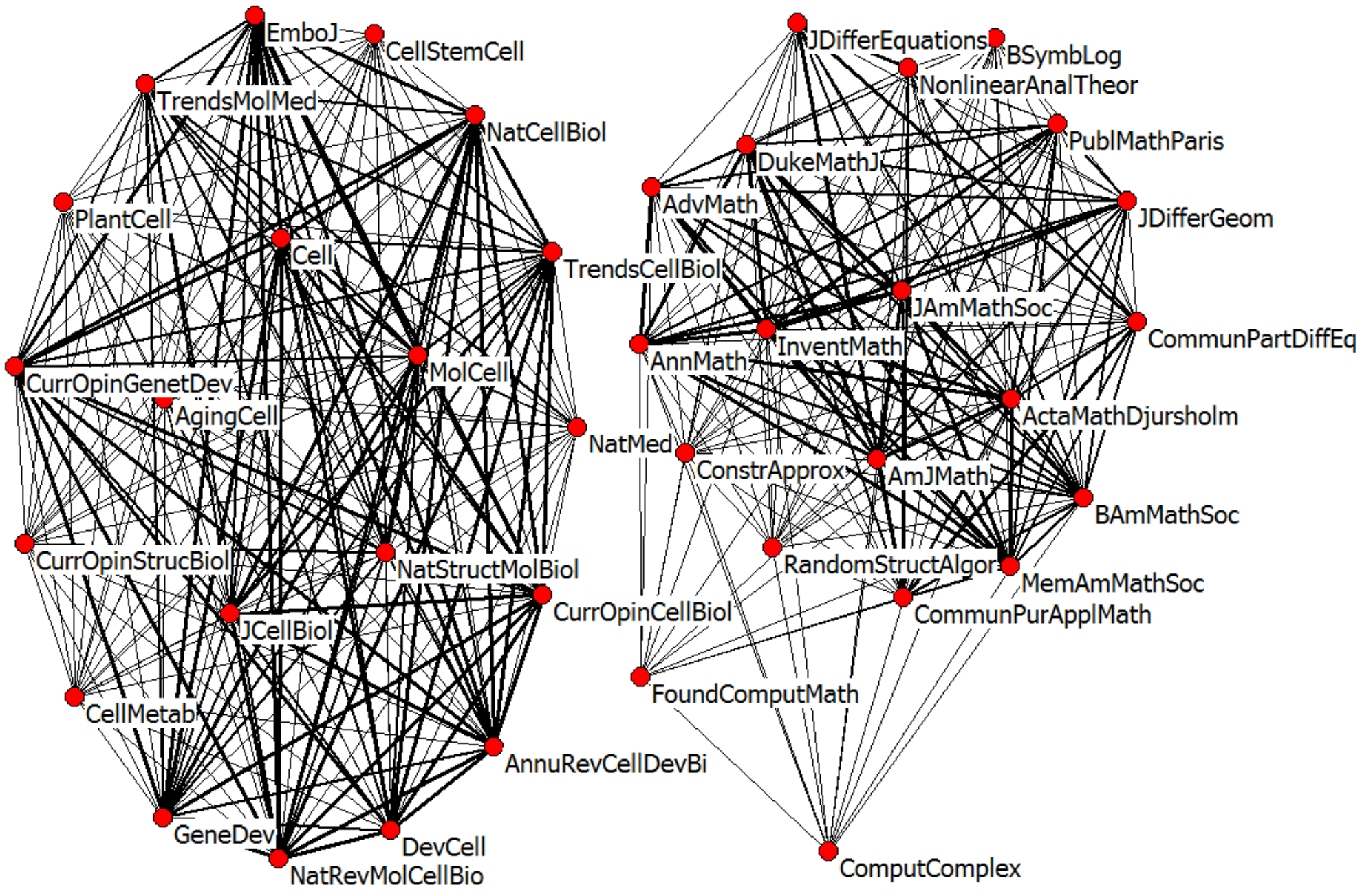

**Figure 1**: Full citation networks among the two sets of 20 journals; no thresholds applied; $N = 40$; layout using Kamada & Kawai (1989) in Pajek.

Figure 1 shows that there is no citation traffic between these two groups of journals when 2008 is used as the publication year citing. Thus, these two groups are fully discrete. (When the map is restricted to references to 2006 and 2007 only, *Computational Complexity* is no longer connected to the mathematics group.) Can this distinction be retrieved by testing the fractionally counted numerators of the quasi-IFs of the 40 journals using a relevant *post-hoc* test?

Among the (2 x 20 =) 40 journals 65,223 references were exchanged in 2008 to the volumes of 2006 and 2007. Between each two citation patterns of these 40 journals, one can test the differences for their statistical significance with ANOVA. Since the variances are again not homogeneous (Levene's test), we use the same Dunnett's C as the *post-hoc* test on the (40 * 39)/ 2 = 780 possible pairwise comparisons.

If two journals are not significantly different in terms of their fractionated citation patterns, they will be considered as belonging to the same group. Figure 2 shows the results for using these two groups of journals—with the black and white colors of the nodes indicating the *a priori* group assignment to mathematics or cellular biology—using Pajek and a spring embedded algorithm (Kamada & Kawai, 1989) for the visualization.

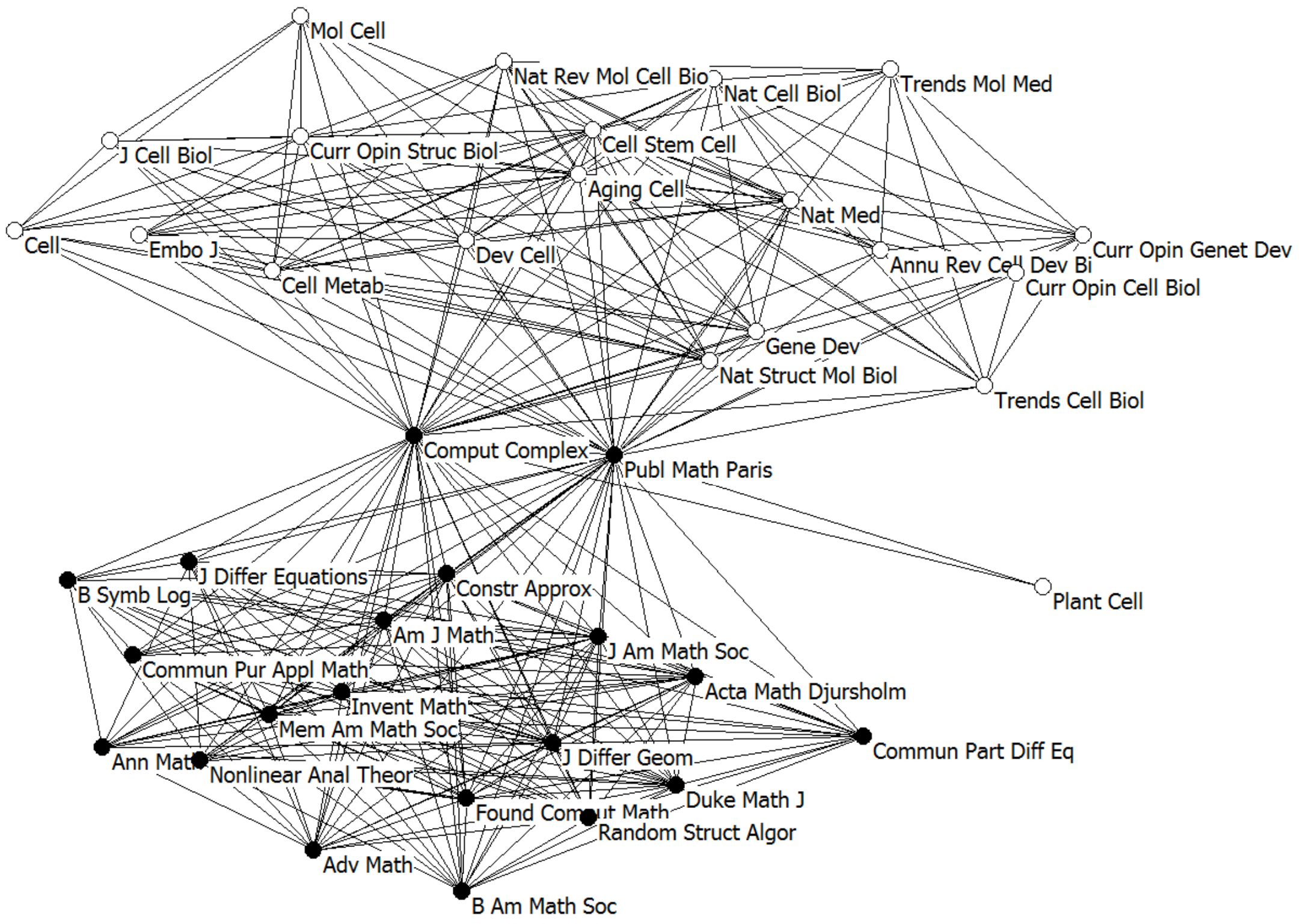

**Figure 2.** Dunnett's C test on fractionally counted citation impacts (2006 and 2007) for two groups of journals.

Journals are linked in the graph (Figure 2) when these statistics are *not* significantly different—in other words, the journals can statistically be considered as a group—in terms of their fractional citation patterns (being cited in 2008). Although these results are motivating on visual inspection, they are not completely convincing. The journal *Plant Cell* is set apart—as it perhaps should be—but its relationships to the mathematics journals *Computational Complexity* and *Publications Mathématiques de l'IHÉS (Paris)* are unexpected. The patterns in these latter two journals deviate from their group (of mathematics journals) and accord also with other groupings.

One measure of the quality of the classification can be found in the density of the two networks depicted in Figure 2. Table 6 provides the densities and average degrees for both the fractionally counted and integer counted sets and subsets; both for the numerator of the IF and the total cites.

| | | Complete set *N* = 40 | | Cell Biology *N* = 20 | | Mathematics *N* = 20 | | Between Partitions |
|---|---|---|---|---|---|---|---|---|
| | | Density | average degree | density | average degree | density | average degree | Density |
| IF Numerator | Fractionally counted | 0.41 | 31.6 | 0.53 | 20 | 0.93 | 35.2 | 0.11 |
| | Integer counted | 0.50 | 39.2 | 0.25 | 7.8 | 0.88 | 33.4 | 0.49 |
| Total Cites | Fractionally counted | 0.28 | 22.0 | 0.14 | 5.4 | 0.57 | 21.6 | 0.22 |
| | Integer counted | 0.24 | 18.8 | 0.09 | 3.6 | 0.37 | 14.2 | 0.26 |

**Table 6**: Densities and average degrees of the top-20 journals in the ISI Subject Categories of *Mathematics* and *Cell Biology* when networked in terms of the significance of the differences in relevant citation distributions.

In accordance with the results of visual inspection of Figure 2, one can observe in Table 6 that the density in the subset of mathematics journals is almost 100% (0.93). On average these journals maintain 35.2 (mutual) relations among the 20 journals. In contrast, however, the density for the group of 20 journals classified as cell biology is 53%. The citation patterns of these journals are significantly different from approximately half of the other journals of this set.

If the same exercise is performed using integer counting, the effects on the mathematics set are not large (– 5%), but the number of links within the group of journals *a priori* classified as cell biology is now only 25%. Furthermore, the number of links between the two partitions increases more than four times to 49%. Thus, the number of misclassifications outside the mathematics group increases significantly using integer counting.

We repeated the same exercise using not only the citations to the two previous years—that is, the numerators of the IFs—but the total cites to these 40 journals: 270,595 references are provided in 2008 to papers in these 40 journals. The larger size of this sample (415%) and the inclusion of citations to all previous years might make it easier to distinguish the two sets, but it did not! The two lower rows in Table 6 show that in this case the density of the relations among the mathematics journals also decreases considerably, reaching a low of only 37% when integer counting is used.

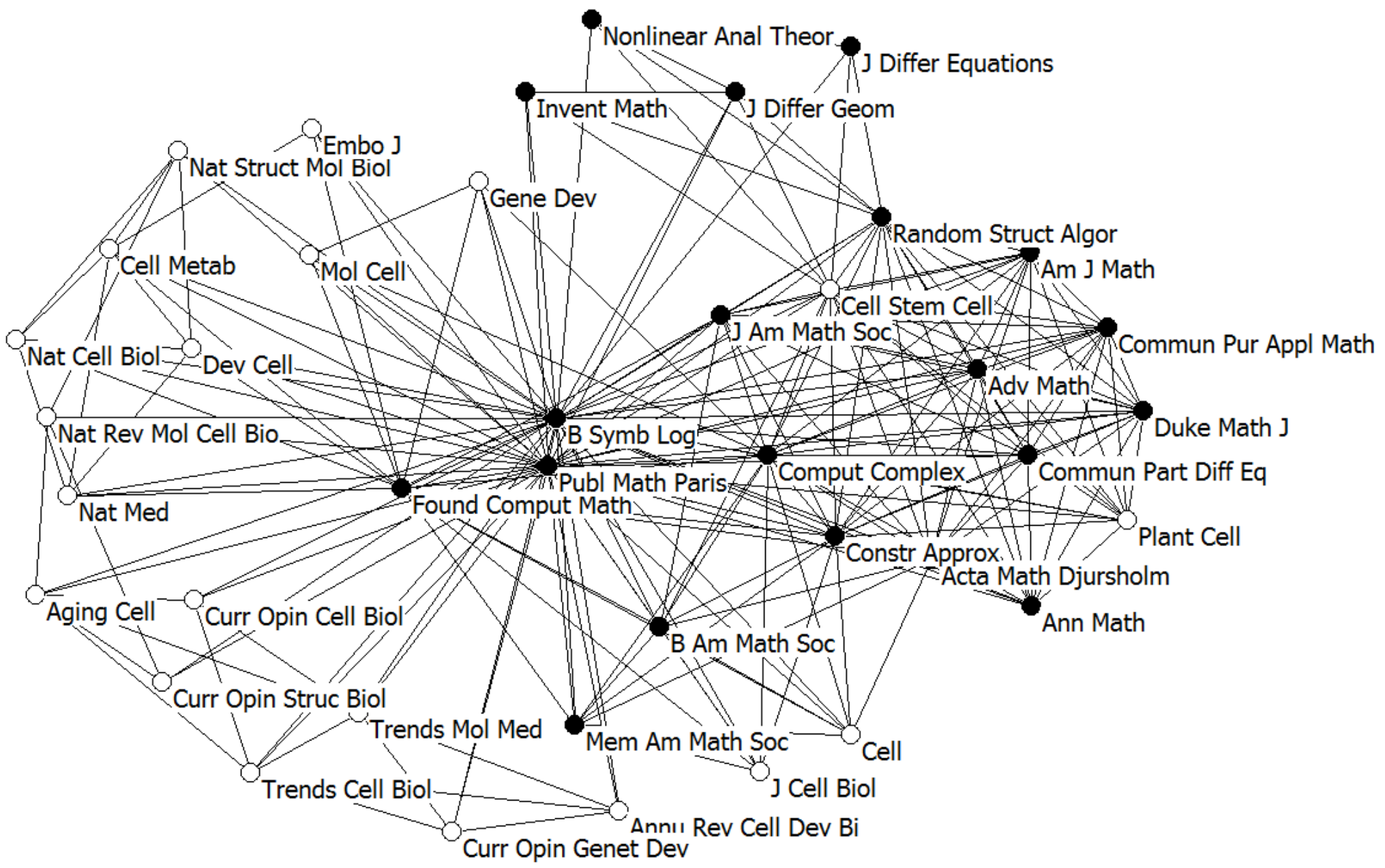

**Figure 3**: Mapping based on fractional counting of total cites in 2008; $N = 40$; Dunnett's C test; visualization in Pajek using Kamada & Kawai (1989).

Figure 3 shows the results of mapping the relations that are not significantly different in terms of their fractionated citation distributions, but using the full set of total cites (instead of only the references to 2006 and 2007). Some journals (e.g., *Cell Stem Cell*—a relatively new journal—but also *Cell*) are now misplaced within the mathematics set.

In summary, the relations at the research front as indicated by the fractionated IF—that is, using only the last two years—are more distinctive than the total cites (that is, taking a longer time span into account). Similarly, a representation based on integer counting in the numerator of the IF (not shown) confirmed that this methodology can only be used for this purpose on the fractionally counted numerator of the quasi-IF.

Even then, the classification in terms of the significance of relations is not reliable. For example, within the group of the 20 mathematics journals, the fractionated citation pattern of the *Journal of Differential Equations* is tested as one of the few significantly different from *Communications in Partial Differential Equations.* In any more standard journal mapping techniques (such as shown in Figure 1), these two journals are visible as strongly related. In our opinion, this result refutes the idea that this test on fractionated citation patterns can reliably be used to sort cognitive differences among journals in terms of fields and specialties.

In summary, the distinction between sets of journals representing different disciplines and specialties cannot be performed using the fractional citation characteristics of the distributions. There remains the question of whether the quasi-IFs based on fractional counting correct sufficiently in a statistical sense for the different citation potentials among the broader disciplines. As noted in the methods section above, we used the thirteen broadly defined fields of the *Science & Engineering Indicators* (2010) for this specification using a variance-component model.

**Variance-component model**

The thirteen fields (NSB, 2010, at p. 5-30 and Appendix Table 5-24) provide the level-2 clusters, and the (quasi-) IFs of the journals are the level-1 units for this test. The research question is whether the differences among fields of sciences (that is, the between-field variance) can be reduced significantly by the normalization of the numerators of the IFs in terms of fractional citation counts. For reasons specified above, we defined additionally a model using the fractional c/p ratios as the dependent variable.

The results of the model estimations are presented in Table 7. We calculated four models (M1 to M4)—each using a different method of measuring journal impact: ISI-IFs 2008, quasi-IFs based on integer counting, quasi-IFs based on fractional counting, and fractionated c/p ratios for 2008. The models assume the intercept as a fixed effect and the variance of the intercepts across fields as a random effect. There are 3,923 (M1 to M3) or 3,869 (M4) IFs of journals, respectively, that are clustered within the 13 fields.[18]

| | M1: ISI-IF 2008 | M2: IF (integer counting) | M3: IF (fractional counting) | M4: Fractionated c/p ratio 2008 |
|---|---|---|---|---|
| Term | Estimate (S.E.) | Estimate (S.E.) | Estimate (S.E.) | Estimate (S.E.) |
| *Fixed effect* | | | | |
| Intercept | .67 (.11)* | .02 (.20) | -1.28 (.10)* | -.75 (.19)* |
| *Random effect* | | | | |
| Level 2 | .15 (.06)* | .48 (.21)* | **.09 (.05)** | .28 (.15) |
| $N_{journal}$ | 3923 | 3923 | 3923 | 3869 |
| $N_{fields\ (clusters)}$ | 13 | 13 | 13 | 13 |

* $p < .05$

**Table 7:** Results of four two-level random-intercept Poisson models

Our assumption is that the level-2 (between-field) variance is reduced (or near zero) by using the IF based on fractional counting (M3) or the fractionated c/p ratio (M4), respectively, compared to the IF based on integer counting (M2). A reduction of this variance coefficient to close to zero would indicate that systematic field differences no longer play a role. The model for the ISI-IF (M1) is additionally included in Table 7; however, only the models M2 to M4 can be compared directly, because for these models the values for each journal are calculated on the basis of the same citation impact data.

[18] 54 journals contained in the CD-Rom version of the *SCI* are not provided with a number of issues in the *JCR* 2008.

The results in Table 7 show that the variance component in the models M1 and M2 are statistically significant. In other words, both sets of data contain statistically significant differences between the fields. However, the variance component is not statistically significant in the models M3 and M4: field differences are no longer significant when the comparison is made in terms of fractionally counted citations. In the comparison of models M3 and M4 with model M2, the level 2-variance component is reduced by ((.48 – .09)/.48)*100) = 81% in model M3 and by ((.48 – .28)/.48)*100) = 42% in model M4.

In summary, the largest reduction of the in-between group variance is associated with model M3; in this case, the in-between group variance component is close to zero. This result provides a very good validation of our assumption: field differences in IFs are significantly reduced—to near zero—when the IFs are based on fractional counting. Using the longer time window as in the case of the c/p ratios does not improve on this result. In other words, these results point out that the quasi-IF based on fractional counting of the citations provides a solution for the construction of an IF where journals can be compared across broadly defined fields of science.

## Conclusions

Further testing using other sets (e.g., the multidisciplinary one mentioned above) confirmed our conclusion that differences in citation potentials cannot be used to distinguish among fields of science statistically. While citation potentials differ among fields of science and, therefore, one can normalize the IFs using fractionated citation counts, this reasoning cannot be reversed. First, other factors obviously play a role such as the differences among document types (e.g., reviews versus research articles and conference proceedings) which are also unevenly distributed among fields of science. Relevant citation windows can also be expected to vary both among fields and over time. In addition to citation behavior, publication behavior varies among fields of science. In other words, the intellectual organization can be expected to affect the textual organization in ways that are different from the statistical expectations based on regularities in the observable distributions (Leydesdorff & Bensman, 2006; Milojević, 2010).

We thought it nevertheless useful to perform the above exercise. The delineation among fields of science—and at the next-lower level, specialties—has hitherto remained an unsolved problem in bibliometrics because these delineations are fuzzy at each moment of time (e.g., each year) but developing dynamically over time. It would have been convenient to have a statistical measure to compare journals with each other on the basis of the citation distributions contained in them at the (citing) article level. We found that a focus on the last two years—following Garfield's (1972) suggestion to follow Martyn &

Gilchrist's (1968) delineation of a "research front"—worked better than including the complete historical record (that is, "total cites"). This conclusion accords with Althouse *et al.*'s (2009) observation that over time citation inflation affects variation more than differences among fields.

Our negative conclusion with respect to the statistical delineation among fields of science does not devalue the correction to the IFs that can be made by using fractional citation counts instead of integer ones. One major source of variance could be removed in this way. In addition to this static variance—in each yearly JCR—the dynamic variance can be removed by using total citations (that is, the complete citation window) instead of the window of the last two years. However, this model did not improve on the regression model using fractional counting for the last two years. The remaining source of variance perhaps could be found in different portfolios among disciplines in terms of document types (reviews, proceedings papers, articles, and letters).[19]

Moed (2010) proposed omitting letters when developing the SNIP indicator arguing that letters and brief communications inflate the representation of the research front by using more references to the last few years. Similarly, one could argue against using reviews because they may deflate the citation potential based on the most recent years (Leydesdorff, 2008, at p. 280, Figure 3). Review articles, however, are currently defined by Thomson Reuters among others as articles that contain 100 or more references.[20] One

[19] The ISI (Thomson Reuters) decided to divide the category of "articles" into "articles" and "proceedings papers" as of October 2008.
[20] "In the *JCR* system any article containing more than 100 references is coded as a review. Articles in 'review' sections of research or clinical journals are also coded as reviews, as are articles whose titles

could focus exclusively on articles and proceedings papers, but in this study we wished to compare the effects of fractionation directly with the ISI-IF which is based on integer counting of the citations of all "citable items".

In other words, differences in publication behavior—perhaps to be distinguished from citation behavior—can be expected to provide yet another source of variance (Ulf Sandström, *personal communication*, March 5, 2010). Furthermore, the fuzziness of the delineations may be generated by creative scholars who are able to move and cite "interdisciplinarily" among fields and specialties (Edge & Mulkay, 1976) and thus provide variation to the intellectual organization of the textual structures among journals (Lucio-Arias & Leydesdorff, 2009). However, movements among broadly defined fields of science are exceptional and less likely to affect the statistics.

**Acknowledgement**

We are grateful to Tobias Opthof, Ludo Waltman, and two anonymous referees for communications about previous drafts.

---

contain the word 'review' or 'overview.'" At http://thomsonreuters.com/products_services/science/free/essays/impact_factor/ (retrieved June 18, 2010). See for problems with these delineations among document types in the Scopus database Leydesdorff & Opthof (2010c).